\documentclass{article}
\usepackage{epsfig}
\usepackage{amssymb}
\newcommand{\bfr}{\begin{flushright}}
\newcommand{\efr}{\end{flushright}}
 
\begin{document}
\title{Quantum fluctuation of stress tensor and black holes in three
dimensions
}
\author{
Kiyoshi Shiraishi\\
Akita Junior College, Shimokitade-Sakura, Akita-shi, Akita 010, Japan\\
and\\
Takuya Maki\\
Department of Physics, Tokyo Metropolitan University,\\
Minami-ohsawa, Hachioji-shi, Tokyo 192-03, Japan\\
}
\date{Phys. Rev. {\bf D49} (1994) pp. 5286-5294 
}
\maketitle
\begin{abstract}
The quantum stress tensor near a three-dimensional black hole is
studied for a conformally coupled scalar field. The back reaction to
the metric is also investigated.
\end{abstract}

\section{INTRODUCTION}
There are several styles in studying the interconnection
between gravity and quantum field theory for the
present.%
\footnote{To view the range of the research in quantum
gravity, see Ref. \cite{1}.}
One of the orthodox approaches to the subject
is to investigate the nature of quantum fields in curved
space-times \cite{2}. For example, quantum field theory
around black holes, where the ultimately strong gravitational
field exists, has been examined by many authors \cite{3}.

Recently, the black hole solution to the three dimensional
Einstein equations with a negative cosmological
constant has been found \cite{4} and various aspects of
the black hole (BH) have been analyzed by many authors
\cite{4,5,6}. Because of the reduction of the dynamical degrees
of freedom, the study of quantum field theory near
the three-dimensional BH (3DBH) may yield important
and essential knowledge on the interface between quantum
field theory and gravity.

The aim of this paper is to examine the vacuum polarization
of the stress-energy tensor for conformally coupled
massless scalar fields in the nonrotating BH background
in three dimensions and discuss the back reaction
to the metric due to the quantum stress tensor. In our
previous paper, we calculated the vacuum expectation
value of $\langle\varphi^2\rangle$ from the propagator obtained by the
mode sum method \cite{6}. Here we derive a more general expression
for the propagator by using knowledge of field
theory in anti-de Sitter space. Due to the structure of the
spacetime of our interest, we have one parameter which
comes from the boundary condition. Since there seems to
be no condition that determines the parameters as long as
we treat only a scalar field, we keep this as a free parameter
in calculations of the propagator and other quantities.
Those derived in our previous paper can be understood as
ones which satisfy a certain special boundary condition.

One of the interesting problems is how the quantum fluctuations
do affect the metric, i.e., of the back reaction to
the metric. We study the problem by evaluating the increase
or decrease of the mass and of radial accelerations
of a test particle due to the quantum fluctuations of the
stress tensor. We show that there is a certain critical radius
where the correction to the acceleration changes
their signature.

In Sec.~II, we obtain the propagator for a conformally
coupled massless scalar field in the 3DBH space-time. In
Sec.~III, the expectation value of $\langle\varphi^2\rangle$ and the
stress tensor for the scalar field are calculated from the propagator.
We consider the back reaction of the quantum
effects to the space-time metric around the 3DBH in Sec.~IV. Discussion
is given in Sec.~V.

\section{PROPAGATOR
FOR CONFORMALLY COUPLED SCALAR FIELDS}
The action for the three-dimensional gravity considered
by the authors in Ref.~\cite{4} is given by
\begin{equation}
S=\int d^3x\frac{\sqrt{-g}}{2\pi} (R+2\lambda)+(\mbox{surface term})\,,
\label{2.1}
\end{equation}
where $R$ is the scalar curvature and $\lambda$ the cosmological
constant. Our conventions follow the textbook \cite{7}. Applying
the variational principle to the action (\ref{2.1}), one
can derive the classical Einstein equations
\begin{equation}
R_{\mu\nu}-\frac{1}{2}Rg_{\mu\nu}=\lambda g_{\mu\nu}\,, 
\label{2.2}
\end{equation}
provided that there is no matter field.

The authors of Ref.~\cite{4} have found the BH solution to
the Einstein equation (\ref{2.2}), which can be written by the
metric
\begin{equation}
ds^2=-(\lambda r^2-M) dt^2 +\frac{dr^2}{\lambda r^2-M}+r^2 d\theta^2\,,
\label{2.3}
\end{equation}
for the nonrotating case. $M$ is the mass of the 3DBH \cite{4}.
The horizon length is $r_H=(M/\lambda)^{1/2}$.

A new coordinate $\rho$ defined by
\begin{equation}
r=r_H\sec\rho\qquad (0\le\rho\le\pi/2)\,, 
\label{2.4}
\end{equation}
makes the metric very simple to handle; then the metric
becomes
\begin{equation}
ds^2=\lambda^{-1}(\sec\rho)^2(-M\lambda\sin^2\rho\, dt^2 +d\rho^2
 +M d\theta^2)\,.
\label{2.5}
\end{equation}

Further substituting the time coordinate $t$ by $-i\tau$, we
obtain the Euclidean metric
\begin{equation}
ds^2_E=\lambda^{-1}(\sec\rho)^2(d\rho^2+\kappa^2\sin^2\rho\, d\tau^2
+M d\theta^2)\,.
\label{2.6}
\end{equation}
where we set $\kappa^2=M\lambda$.

We consider a scalar field in this background spacetime.
We consider a conformally coupled, massless scalar
field in three dimensions in this paper. The wave equation
for the scalar field is
\begin{equation}
\Box\varphi-\frac{1}{8}R\varphi=0\,,
\label{2.7}
\end{equation}
where the covariant divergence is defined in terms of the
background metric (\ref{2.6}). The propagator $G_E$ for this scalar
field in the Euclidean geometry is the solution to the
following equation:
\begin{equation}
\left(\Box-\frac{1}{8}R\right)G_E(x, x')=-\frac{1}{\sqrt{g}}\delta(x,
x')\,. 
\label{2.8}
\end{equation}

In our previous paper (Ref.~\cite{6}) we have taken the
mode sum method to obtain an exact form of the propagator. In this
paper, we derive the exact form of the propagator by using the
knowledge of the field theory in an anti-de Sitter space
\cite{8,9,10,11}.

If one replaces $\sqrt{M}\theta$ in the metric (\ref{2.6}) by a new
angular coordinate $\tilde{\theta}$, one finds that the metric seems to
be the one for the Euclideanized anti-de Sitter space (EAdS).
Thus, we find that the Euclidean 3DBH space-time has
the same local structure as the one of EAdS. Therefore,
we can utilize the knowledge on the field theory in the
anti-de Sitter space. At the same time, we must note,
however, that the global structure of the BH space-time
differs from that of the EAdS; in the 3DBH space-time,
there is a BH horizon but no closed timelike curve \cite{4}.

The most important notion is in the fact that the
3DBH space-time can be regarded as a quotient space of
the anti-de Sitter space. The substitution
$\sqrt{M}\theta\rightarrow\tilde{\theta}$ as previously stated brings
about a ``negative deficit angle,'' because the coordinate
$\tilde{\theta}$ has an unusual periodicity so that
$0\le\tilde{\theta}\le 2\pi\sqrt{M}$. We must guarantee the periodicity
in the angular coordinate even in the Euclideanized space.

The general solution to Eq.~(\ref{2.8}) in the background of
EAdS, whose metric is obtained by the substitution
$\sqrt{M}\theta\rightarrow\tilde{\theta}$ in (\ref{2.6}), can be written
as \cite{8,9,10,11} 
\begin{eqnarray}
& &G_{E}^{EAdS}(\rho, \tau, \tilde{\theta}; \rho', \tau',
\tilde{\theta}')\nonumber \\
& &=\frac{\sqrt{\lambda}(\cos\rho)^{1/2}
(\cos\rho')^{1/2}}{
4\sqrt{2}\pi\sqrt{\cos(\tilde{\theta}-\tilde{\theta}')-\cos\rho
\cos\rho'-\sin\rho\sin\rho'\cos\kappa(\tau-\tau')}}\nonumber \\
& &
-\alpha
\frac{\sqrt{\lambda}(\cos\rho)^{1/2} (\cos\rho')^{1/2}}{
4\sqrt{2}\pi\sqrt{\cos(\tilde{\theta}-\tilde{\theta}')+\cos\rho
\cos\rho'-\sin\rho\sin\rho'\cos\kappa(\tau-\tau')}}\,,
\label{2.9}
\end{eqnarray}
where $\alpha$ is a constant. The second term on the right-hand side of
Eq.~(\ref{2.9}) exhibits a singular point outside the physical region.
We do not specify the value of a here by the consideration of symmetry
group or by the causality usually suggested in the field theory in an
anti-de Sitter space because of the difference in the global structure
as previously mentioned.
We adopt the ``mixed'' boundary condition including a parameter a
throughout this paper.%
\footnote{A possibility that the consideration of supersymmetry or
other physics may prefer a certain boundary condition is left to be
investigated for another opportunity.}

To construct the propagator in the 3DBH
space-time from (\ref{2.9}), we must take the periodicity in the angular
coordinate into the propagator. The construction can be just carried
out by means of the image method \cite{12}. We show the result in the
following form: 
\begin{eqnarray}
& &G_{E}^{BH}(\rho, \tau, {\theta}; \rho', \tau',
{\theta}')\nonumber \\
& &=\sum_{k=-\infty}^\infty\frac{\sqrt{\lambda}(\cos\rho)^{1/2}
(\cos\rho')^{1/2}}{
4\sqrt{2}\pi\sqrt{\cosh\sqrt{M}({\theta}-{\theta}'+2\pi
k)-\cos\rho
\cos\rho'-\sin\rho\sin\rho'\cos\kappa(\tau-\tau')}}\nonumber \\
& &
-\alpha\sum_{k=-\infty}^\infty
\frac{\sqrt{\lambda}(\cos\rho)^{1/2} (\cos\rho')^{1/2}}{
4\sqrt{2}\pi\sqrt{\cosh\sqrt{M}({\theta}-{\theta}'+2\pi
k)+\cos\rho
\cos\rho'-\sin\rho\sin\rho'\cos\kappa(\tau-\tau')}}\,.
\label{2.10}
\end{eqnarray}

This result agrees with the expression derived
by the mode sum method
\cite{6} for $\alpha=0$.%
\footnote{In the mode sum method, we can get the propagator for the
``mixed'' boundary condition in this paper by taking the general linear
combination of two types of Legendre functions as the mode function.}

It is remarkable that we have obtained the exact expression for the
propagator valid at an arbitrary distance from the BH in three
dimensions; no exact solution in a closed form has been known for the
BH space-time in the other dimensions except for two dimensions
\cite{13}.

We can calculate the propagator for a twisted scalar field which has
the antiperiodicity in the angular variable \cite{14}:
\begin{equation}
\varphi(\theta+2\pi)=-\varphi(\theta)\,. 
\label{2.11} 
\end{equation}

For the calculation for the
twisted field around the 3DBH can be done similarly to the previous
untwisted case. One can find the propagator for a twisted scalar field
is given by 
\begin{eqnarray}
& &G_{E,twisted}^{BH}(\rho, \tau, {\theta}; \rho', \tau',
{\theta}')\nonumber \\
& &=\sum_{k=-\infty}^\infty\frac{(-1)^k\sqrt{\lambda}(\cos\rho)^{1/2}
(\cos\rho')^{1/2}}{
4\sqrt{2}\pi\sqrt{\cosh\sqrt{M}({\theta}-{\theta}'+2\pi
k)-\cos\rho
\cos\rho'-\sin\rho\sin\rho'\cos\kappa(\tau-\tau')}}\nonumber \\
& &
-\alpha\sum_{k=-\infty}^\infty
\frac{(-1)^k\sqrt{\lambda}(\cos\rho)^{1/2} (\cos\rho')^{1/2}}{
4\sqrt{2}\pi\sqrt{\cosh\sqrt{M}({\theta}-{\theta}'+2\pi
k)+\cos\rho
\cos\rho'-\sin\rho\sin\rho'\cos\kappa(\tau-\tau')}}\,.
\label{2.12}
\end{eqnarray}
In the next section, we will calculate the vacuum expectation
value for $\langle\varphi^2\rangle$ and the stress tensor for the
conformally invariant scalar field in the 3DBH space-time.

\section{QUANTUM STRESS TENSOR
FOR A CONFORMALLY INVARIANT SCALAR FIELD}
Now we calculate the quantum fluctuations of the conformally
coupled scalar field in the BH space-time. The
quantum effects around the 3DBH can be computed from
the propagator (\ref{2.10}) [and for twisted case (\ref{2.12})].

We take the vacuum polarization $\langle\varphi^2\rangle$ as the
coincidence limit of the propagator $G_E(x, x)$ after appropriate
regularization \cite{2,3,6,11,13}. The expectation value for
the stress tensor, on the other hand, can be obtained by
the following coincidence limit with regularization:
\begin{equation}
\langle T^\mu_\nu(x)\rangle=\lim_{x'\rightarrow x}\frac{1}{4}
\left\{\left[3\nabla^\mu\nabla_{\nu'}-\delta^\mu_\nu
\nabla^\sigma\nabla_{\sigma'}
-\nabla^\mu\nabla_{\nu}-\frac{\lambda}{4}\delta^\mu_\nu
\right]G_E^{BH} (x, x')\right\}\,.
\label{3.1}
\end{equation}

To make the computation procedure easy and clear, we
expand the propagator up to second order in terms of the
coordinate intervals. After straightforward calculation,
we find
\begin{eqnarray}
& &G_E^{BH}(\rho, \tau, \theta; \rho', \tau',
\theta')=\frac{1}{4\pi\sqrt{2}\sigma}
-\alpha\frac{\sqrt{\lambda}}{8\pi}\left[1-\frac{\tan^2\rho
(\kappa\Delta\tau)^2}{8}-\frac{M(\Delta\theta)^2}{8\cos^2\rho}
-\frac{(\Delta\rho)^2}{8\cos^2\rho}\right]\nonumber \\
&
&\quad+\sum_{k=1}^\infty\left[\frac{\sqrt{\lambda}\cos[(\rho+\rho')/2]
}{2\sqrt{2}\pi\sqrt{\cosh 2\pi\sqrt{M}k-1}}-\alpha
\frac{\sqrt{\lambda}\cos[(\rho+\rho')/2]
}{2\sqrt{2}\pi\sqrt{\cosh
2\pi\sqrt{M}k+\cos(\rho+\rho')}}\right]\nonumber \\
& &+(\kappa\Delta\tau)^2\sum_{k=1}^\infty
\left[\frac{\sqrt{\lambda}\cos\rho(\cos 2\rho-1)
}{16\sqrt{2}\pi({\cosh 2\pi\sqrt{M}k-1})^{3/2}}-\alpha
\frac{\sqrt{\lambda}\cos\rho(\cos 2\rho-1)
}{16\sqrt{2}\pi({\cosh
2\pi\sqrt{M}k+\cos 2\rho})^{3/2}}\right]\nonumber \\
& &+M(\Delta\theta)^2\sum_{k=1}^\infty
\left[\frac{\sqrt{\lambda}\cos\rho(\cosh 2\pi\sqrt{M}k+3)
}{16\sqrt{2}\pi({\cosh 2\pi\sqrt{M}k-1})^{3/2}}\right.\nonumber \\
& &\qquad\qquad\qquad\qquad\left.-\alpha
\frac{\sqrt{\lambda}\cos\rho(\cosh^2 2\pi\sqrt{M}k-2\cos 2\rho
\cosh 2\pi\sqrt{M}k-3)
}{16\sqrt{2}\pi({\cosh
2\pi\sqrt{M}k+\cos 2\rho})^{5/2}}\right]\nonumber \\
& &-(\Delta\rho)^2\sum_{k=1}^\infty
\left[\frac{\sqrt{\lambda}\cos\rho(\cosh 2\pi\sqrt{M}k+\cos 2\rho)
}{16\sqrt{2}\pi\cos\rho({\cosh 2\pi\sqrt{M}k-1})^{3/2}}\right.\nonumber
\\ & &\qquad\qquad\qquad\qquad\left.-\alpha
\frac{\sqrt{\lambda}}{16\sqrt{2}\pi\cos\rho({\cosh
2\pi\sqrt{M}k+\cos 2\rho})^{1/2}}\right]+O[(\Delta x)^2]\,,
\label{3.2}
\end{eqnarray}
where $\Delta\tau=\tau-\tau'$, $\Delta\theta=\theta-\theta'$, and
$\Delta\rho=\rho-\rho'$. The divergence in the coincidence limit comes
from the first term on the right-hand side, where
\[
\sigma=\frac{\cosh\sqrt{M}(\theta-\theta')-\cos\rho
\cos\rho'-\sin\rho\sin\rho'\cos\kappa(\tau-\tau')}{\lambda\cos\rho
\cos\rho'}=\frac{s^2}{2}
\]
and $s(x, x')$ is the geodesic distance between $x$ and $x'$.

Usually we need the Schwinger-DeWitt expansion of the propagator with
respect to the powers of the geodesic distance between the two points
to analyze the subtraction of the divergence in the coincidence limit.
Fortunately, there are no relevant terms for subtraction, other than
the term including s, thanks to the lack of the invariant with correct
dimension in the three- (or odd-) dimensional case.%
\footnote{In other words, there is no trace anomaly in
odd dimensions \cite{2}.}
Therefore we can
calculate the regularized value for vacuum values of
$\langle\varphi^2\rangle$ and $\langle T_\nu^\mu\rangle$ from
(\ref{3.2}) after throwing away the first term on the right-hand side.

We find that the vacuum value $\langle\varphi^2\rangle$ in the 3DBH
space-time takes the form as a function of $\rho$: 
\begin{eqnarray}
& &\langle\varphi^2\rangle(\rho)=
-\alpha\frac{\sqrt{\lambda}}{8\pi}\nonumber \\
& &+\sum_{k=1}^\infty\left[
\frac{\sqrt{\lambda}\cos\rho}{
2\sqrt{2}\pi\sqrt{\cosh 2\pi\sqrt{M}k-1}}-\alpha
\frac{\sqrt{\lambda}\cos\rho}{
2\sqrt{2}\pi\sqrt{\cosh 2\pi\sqrt{M}k+\cos 2\rho}}\right]\,.
\label{3.3}
\end{eqnarray}

For $\alpha=0$, the dependence on $\rho$ becomes very simple. In terms
of the original coordinate $r$ in (\ref{2.3}), is inversely proportional
to the radial coordinate $r$. $\langle\varphi^2\rangle$ approaches zero
in the limit of
$r\rightarrow\infty$ if and only if $\alpha=0$. The value of
$\langle\varphi^2\rangle$ at spatial infinity is a constant 
$-\alpha\sqrt{\lambda}/(8\pi)$. 

\begin{figure}[ht]
\begin{center}
\includegraphics[width=5cm]{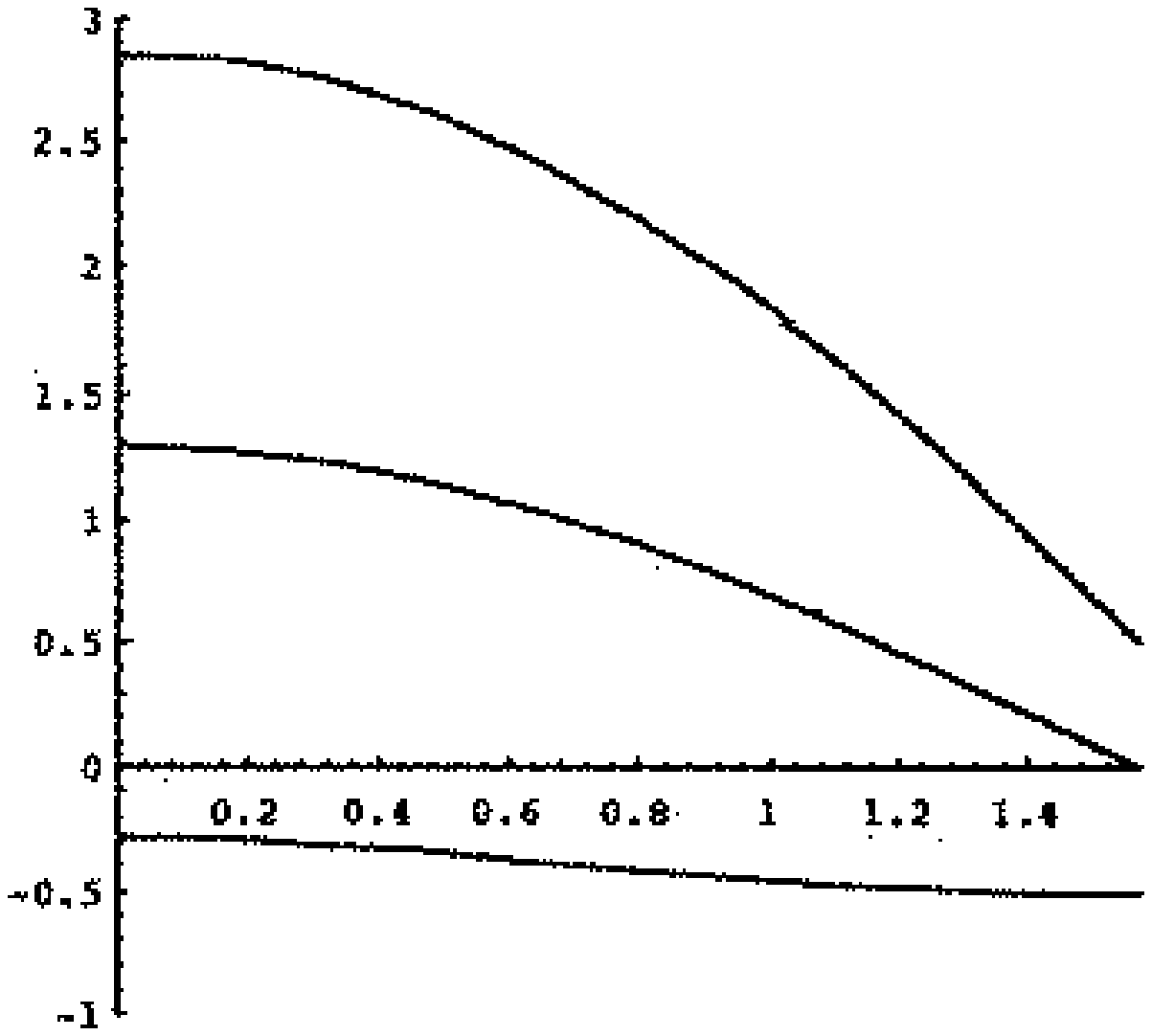}
\caption{The magnitude of the vacuum polarization
$4\pi\langle\varphi^2\rangle(\rho)/\sqrt{\lambda}$ as a function of
$\rho$ when the BH mass $M=\pi^{-2}$. The curves correspond to
$\alpha=-1, 0, 1$ as indicated.}
\label{f1}
\end{center}
\end{figure}

\begin{figure}[ht]
\begin{center}
\includegraphics[width=5cm]{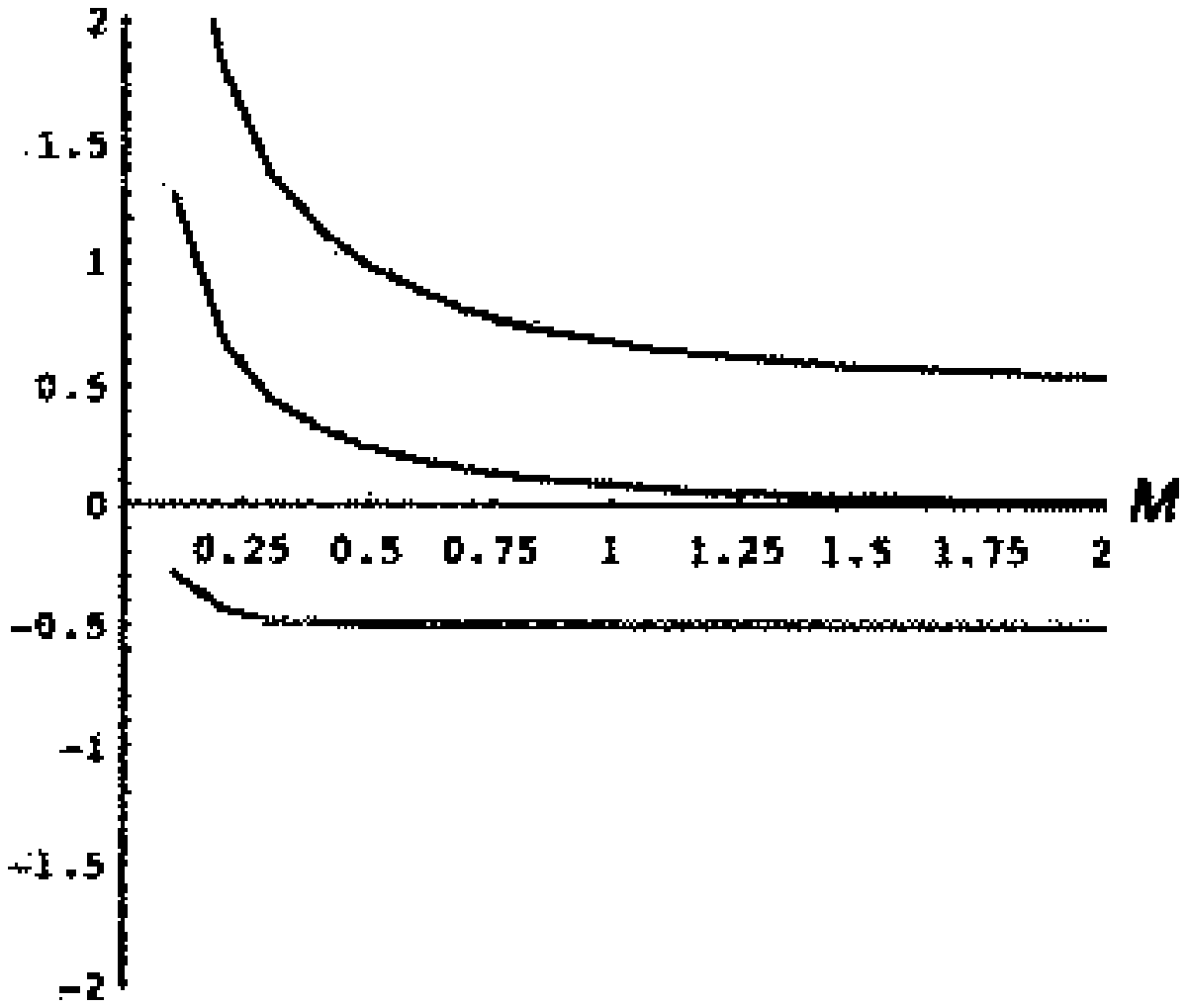}
\caption{The dependence of the vacuum polarization
$4\pi\langle\varphi^2\rangle(\rho)/\sqrt{\lambda}$ on the horizon as a
function of $M$. The curves correspond to $\alpha=-1, 0, 1$ as
indicated.}
\label{f2}
\end{center}
\end{figure}

The value of $4\pi\langle\varphi^2\rangle(\rho)/\sqrt{\lambda}$ is
plotted in Fig.~1 for $\alpha=-1, 0$, and $1$ when the BH mass
$M=\pi^{-2}$. The dependence of $\langle\varphi^2\rangle$ at the
horizon ($\rho=0; r=r_H$) on the BH mass $M$ is shown in Fig.~2. The
value of $\langle\varphi^2\rangle$ at the horizon approaches a constant
$-\alpha\sqrt{\lambda}/(8\pi)$ in the large mass limit. 

\begin{figure}[ht]
\begin{center}
\includegraphics[width=5cm]{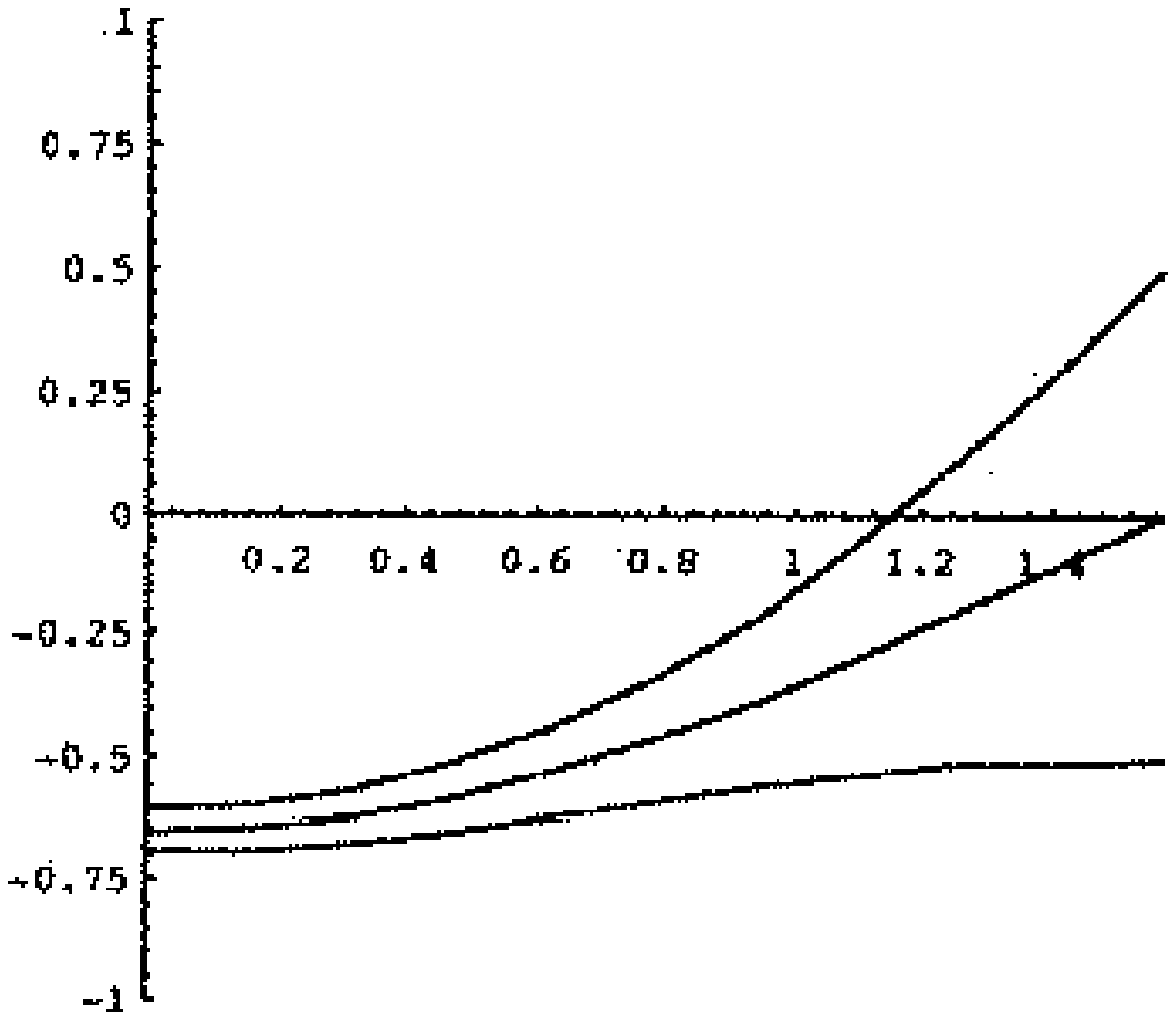}
\caption{The magnitude of the vacuum polarization
$4\pi\langle\varphi^2\rangle_{twisted}(\rho)/\sqrt{\lambda}$ as a
function of $\rho$ when the BH mass $M=\pi^{-2}$. The curves correspond
to $\alpha=-1, 0, 1$ as indicated.}
\label{f3}
\end{center}
\end{figure}

\begin{figure}[ht]
\begin{center}
\includegraphics[width=5cm]{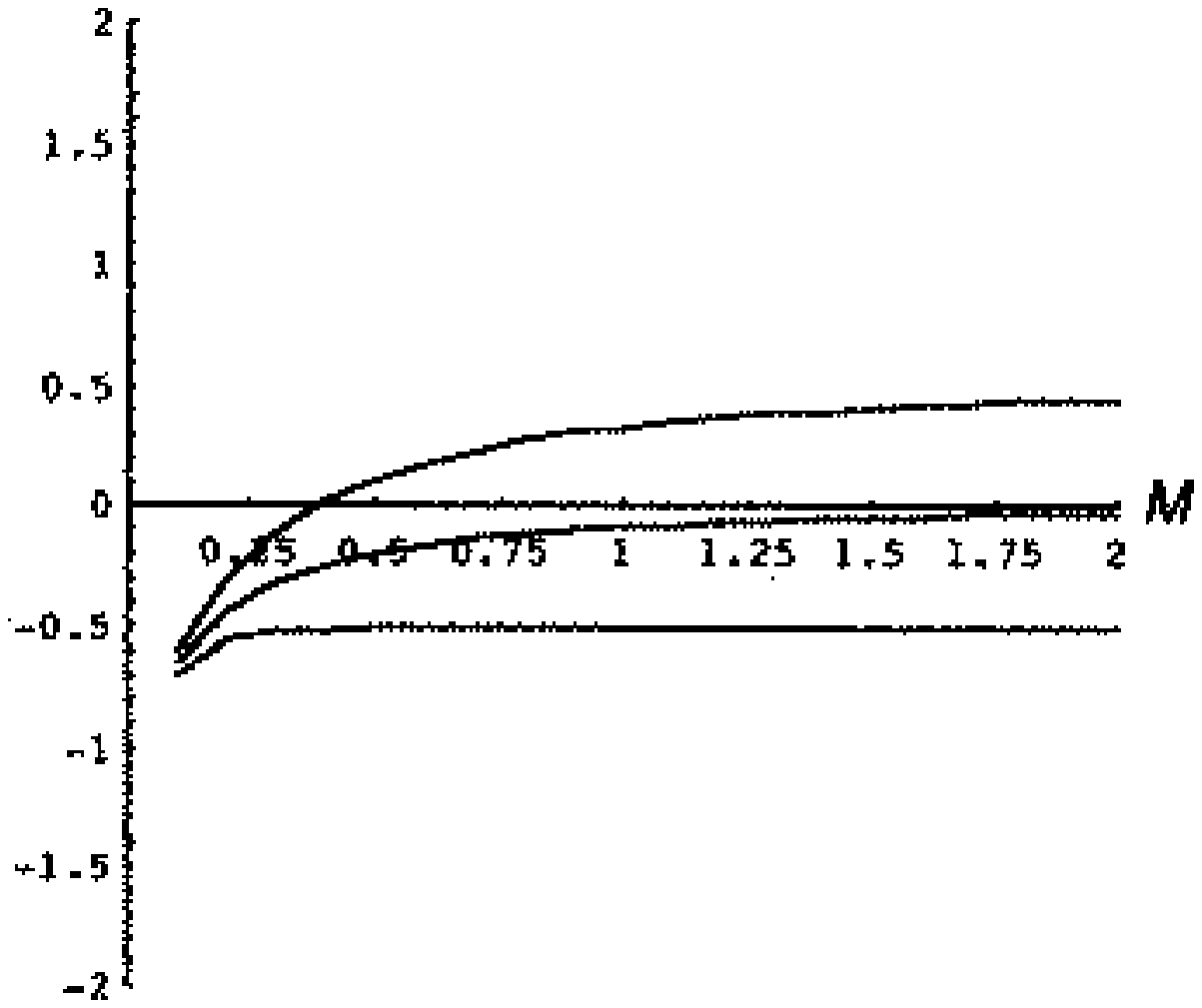}
\caption{The dependence of the vacuum polarization
$4\pi\langle\varphi^2\rangle_{twisted}(\rho)/\sqrt{\lambda}$ on the
horizon as a function of $M$. The curves correspond to $\alpha=-1, 0,
1$ as indicated.}
\label{f4}
\end{center}
\end{figure}

For the twisted scaIar field, we obtain
\begin{eqnarray}
& &\langle\varphi^2\rangle_{twisted}(\rho)=
-\alpha\frac{\sqrt{\lambda}}{8\pi}\nonumber \\
& &+\sum_{k=1}^\infty (-1)^k\left[
\frac{\sqrt{\lambda}\cos\rho}{
2\sqrt{2}\pi\sqrt{\cosh 2\pi\sqrt{M}k-1}}-\alpha
\frac{\sqrt{\lambda}\cos\rho}{
2\sqrt{2}\pi\sqrt{\cosh 2\pi\sqrt{M}k+\cos 2\rho}}\right]\,.
\label{3.4}
\end{eqnarray}
The figures corresponding to Figs.~1 and 2 are exhibited as Figs.~3 and
4 for the twisted scalar field. 

Now we turn to the vacuum stress
tensor. From (\ref{3.1}) and (\ref{3.2}), we find the simple form
\begin{equation}
\langle T^\mu_\nu\rangle(\rho)=F(\rho)\mbox{diag}(1, 1,
-2)+H(\rho)\mbox{diag}(1, 0, -1)\,, 
\label{3.5}
\end{equation}
where
\begin{eqnarray}
& &F(\rho)=\frac{\lambda^{3/2}\cos^3\rho}{16\sqrt{2}\pi}
\sum_{K=1}^\infty\left[
\frac{1}{(\cosh 2\pi\sqrt{M}k-1)^{1/2}}+\frac{4}{(\cosh
2\pi\sqrt{M}k-1)^{3/2}}\right.\nonumber \\
& &
\left.-\alpha\frac{\cosh 2\pi\sqrt{M}k-1}{(\cosh 2\pi\sqrt{M}k+\cos
2\rho)^{3/2}}\right]\,,\label{3.6}\\
& &H(\rho)=-\alpha\frac{3\lambda^{3/2}\sin^2\rho\cos^3\rho}{8\sqrt{2}
\pi}\sum_{K=1}^\infty\frac{\cosh 2\pi\sqrt{M}k-1}{
(\cosh 2\pi\sqrt{M}k+\cos 2\rho)^{5/2}}\,.
\label{3.7}
\end{eqnarray}

Obviously the vacuum stress tensor is traceless. One can also check
the conservation law $\nabla_\mu\langle T^\mu_\nu\rangle=0$ in the
3DBH background. 

For the twisted scalar, we have
\begin{equation}
\langle T^\mu_\nu\rangle_{twisted}(\rho)=F_{twisted}(\rho)\mbox{diag}(1,
1, -2)+H_{twisted}(\rho)\mbox{diag}(1, 0, -1)\,, 
\label{3.8}
\end{equation}
with
\begin{eqnarray}
& &F_{twisted}(\rho)=\frac{\lambda^{3/2}\cos^3\rho}{16\sqrt{2}\pi}
\sum_{K=1}^\infty\left[
\frac{(-1)^k}{(\cosh 2\pi\sqrt{M}k-1)^{1/2}}+\frac{4(-1)^k}{(\cosh
2\pi\sqrt{M}k-1)^{3/2}}\right.\nonumber \\
& &
\left.-\alpha\frac{(-1)^k(\cosh 2\pi\sqrt{M}k-1)}{(\cosh
2\pi\sqrt{M}k+\cos 2\rho)^{3/2}}\right]\,,\label{3.9}\\
&
&H_{twisted}(\rho)=-\alpha\frac{3\lambda^{3/2}\sin^2\rho\cos^3\rho}{8\sqrt{2}
\pi}\sum_{K=1}^\infty\frac{(-1)^k(\cosh 2\pi\sqrt{M}k-1)}{
(\cosh 2\pi\sqrt{M}k+\cos 2\rho)^{5/2}}\,.
\label{3.10}
\end{eqnarray}

The translation to the stress tensor in terms of the
coordinate $(t, r, \theta)$ is trivially performed. In the next section,
we will mainly use the original coordinate system
like (\ref{2.3}).

\begin{figure}[ht]
\begin{center}
\includegraphics[width=5cm]{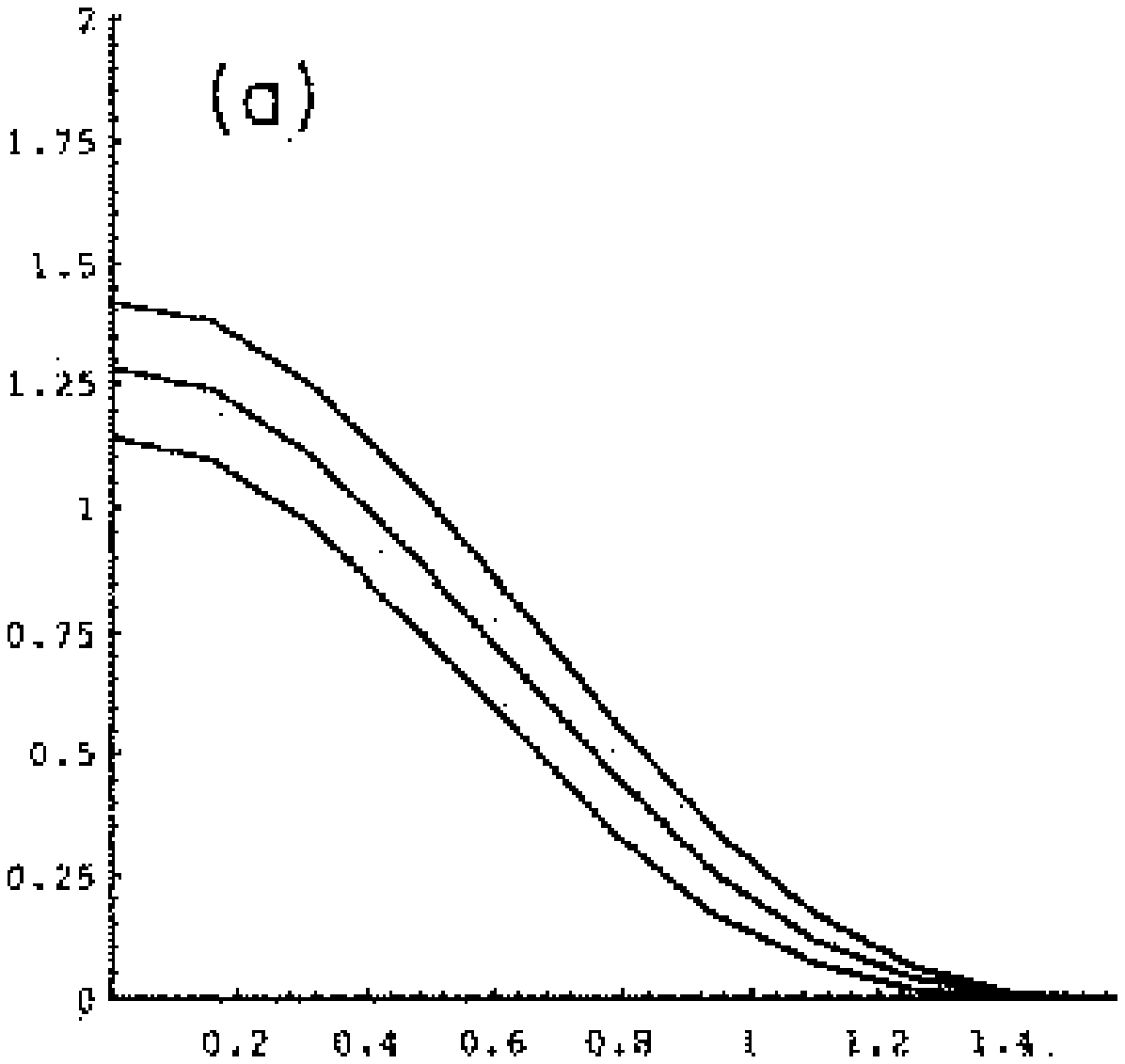}\\
\includegraphics[width=5cm]{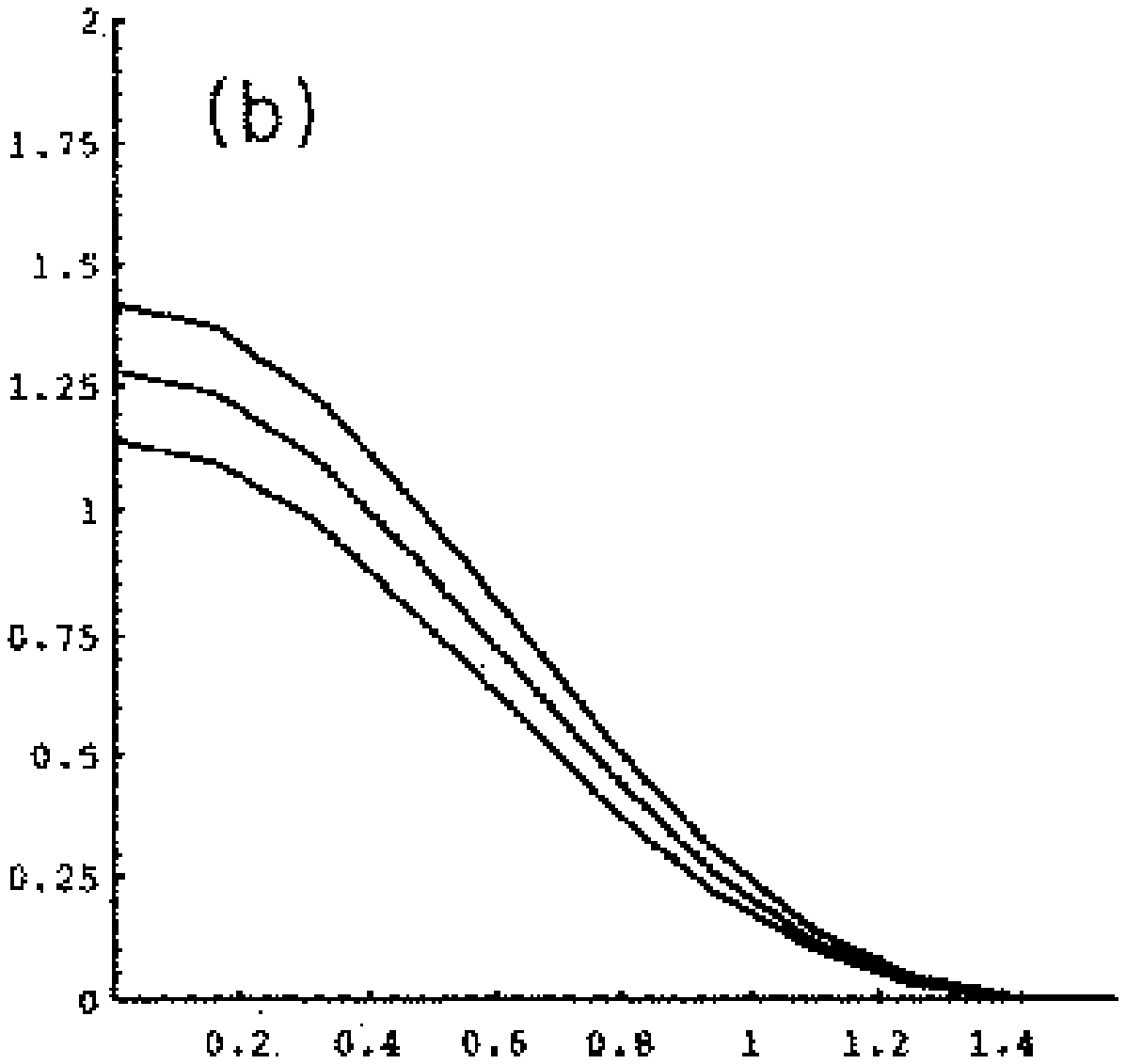}\\
\includegraphics[width=5cm]{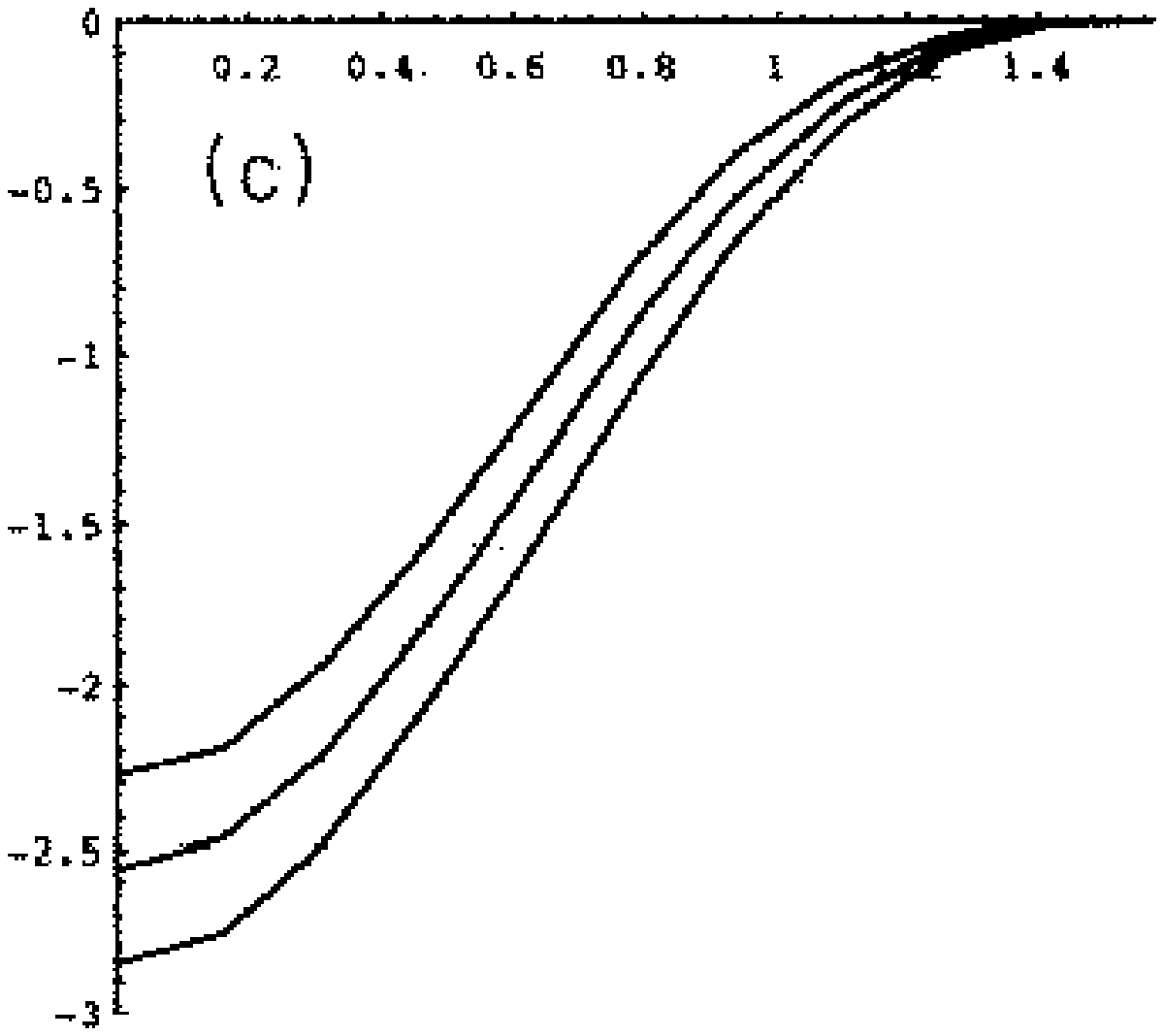}
\caption{The magnitude of the vacuum polarization $16\pi\langle
T^\mu_\nu\rangle(\rho)/\lambda^{3/2}$ as a function of $\rho$ when the
BH mass $M=\pi^{-2}$. Three diagonal components are shown in separate
figures: (a) $(\mu, \nu)=(t, t)$, (b) $(\mu, \nu)=(r, r)$, and (c)
$(\mu, \nu)=(\theta, \theta)$. The curves correspond to $\alpha=-1, 0,
1$ as indicated.}
\label{f5}
\end{center}
\end{figure}
\begin{figure}[ht]
\begin{center}
\includegraphics[width=5cm]{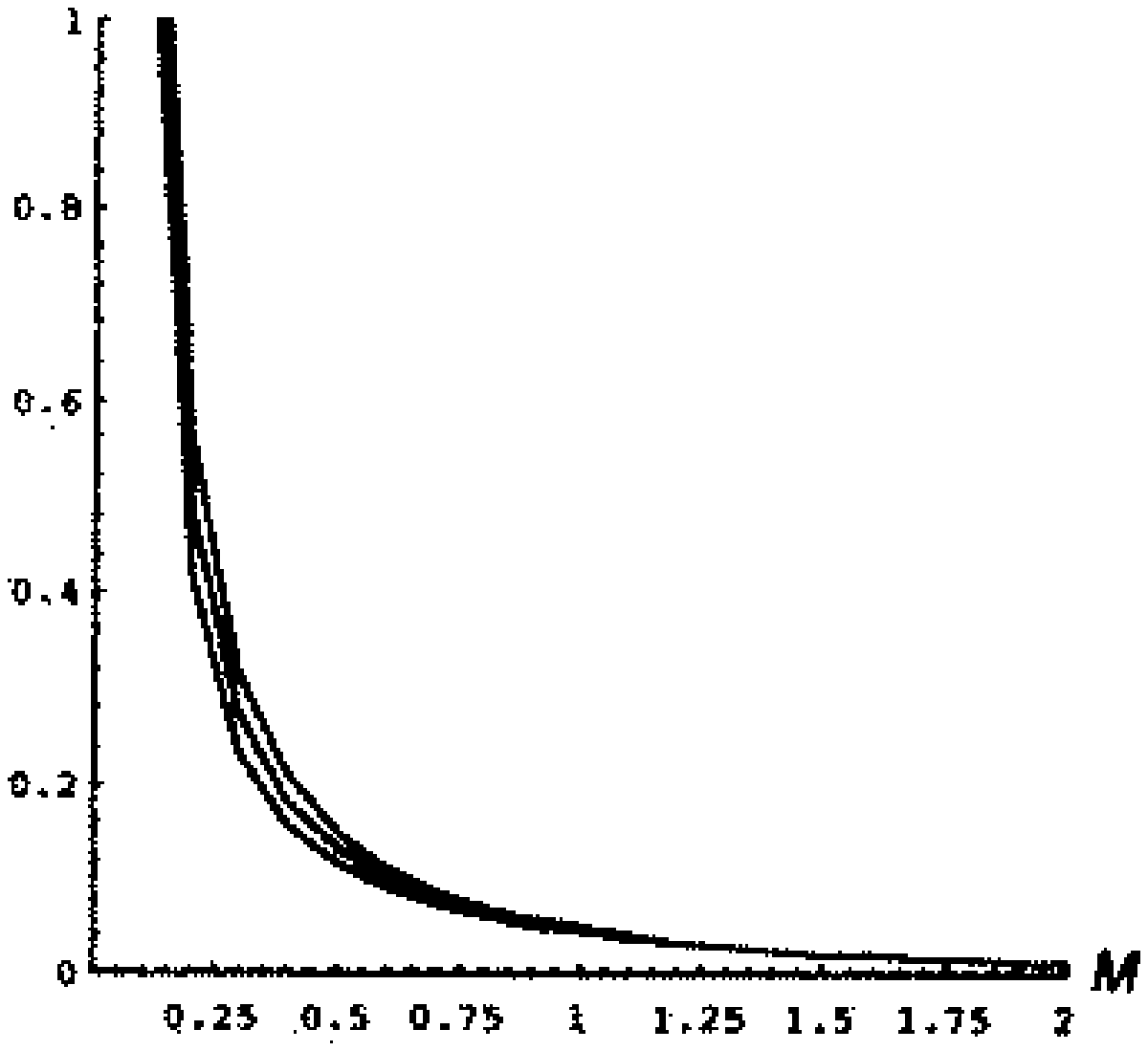}
\caption{The dependence of the vacuum polarization
$16\pi\langle
T^\mu_\nu\rangle(\rho)/\lambda^{3/2}$ on the horizon as a function of
$M$. The curves correspond to $\alpha=-1, 0,
1$ as indicated.}
\label{f6}
\end{center}
\end{figure}
\begin{figure}[ht]
\begin{center}
\includegraphics[width=5cm]{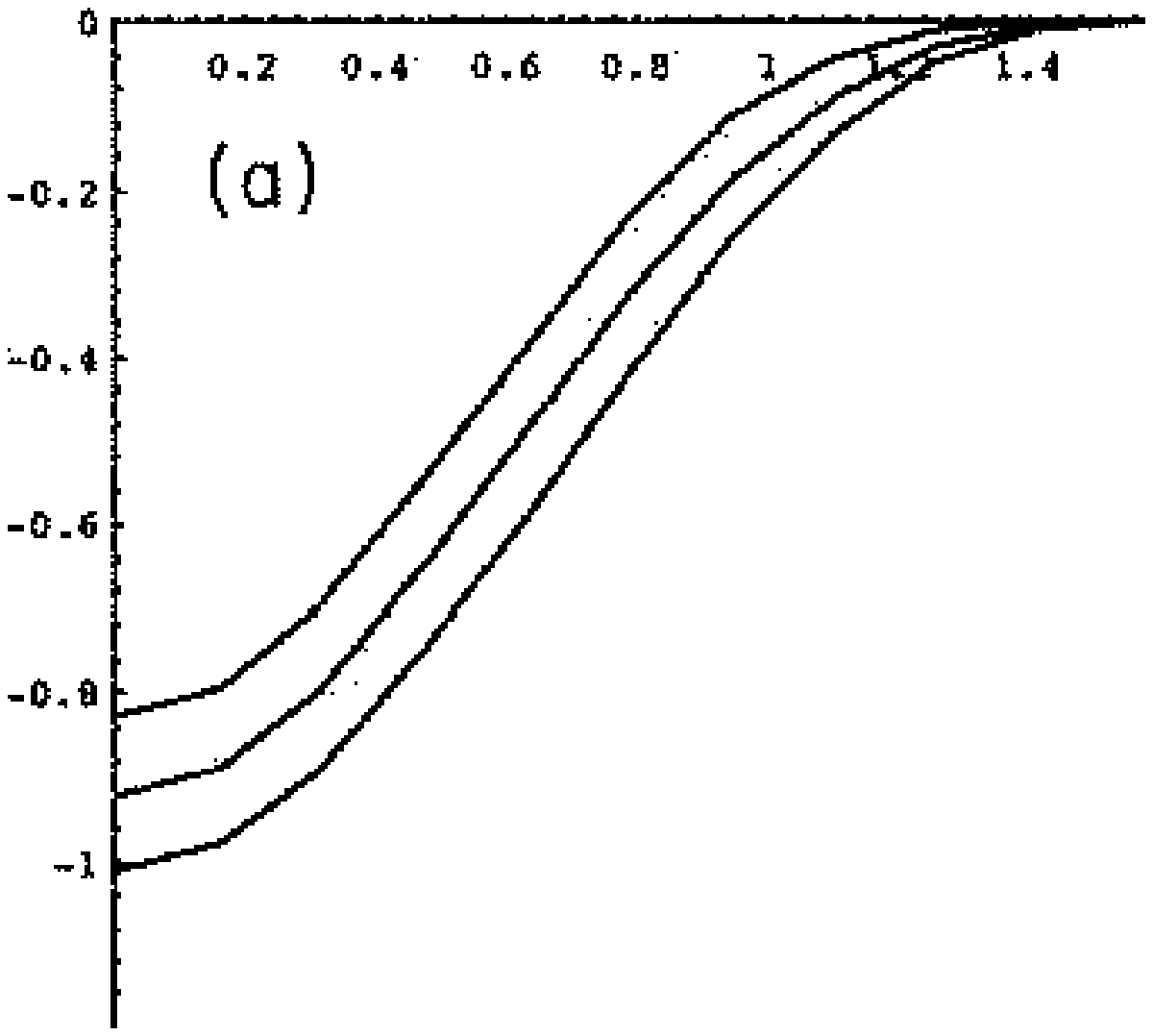}\\
\includegraphics[width=5cm]{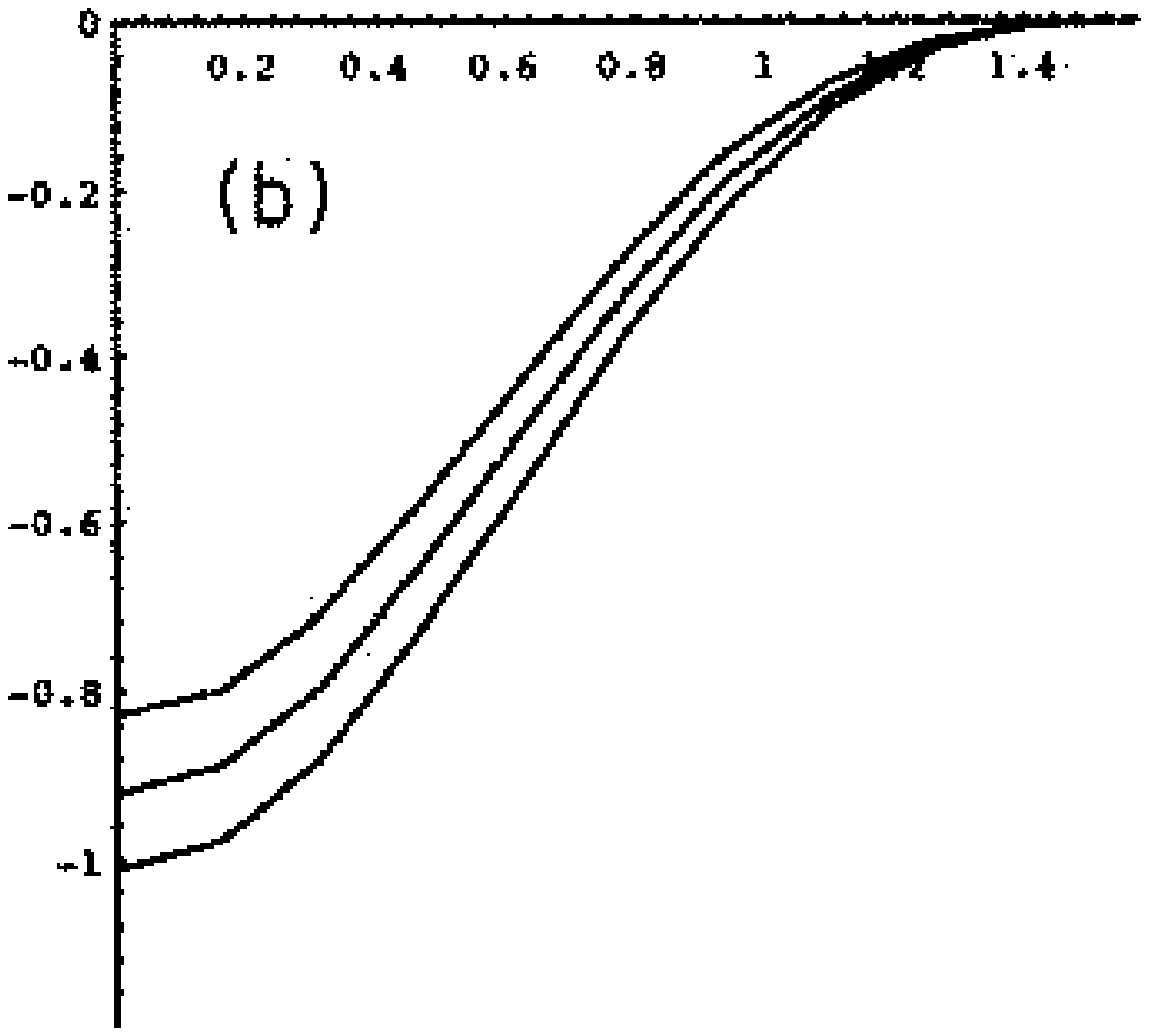}\\
\includegraphics[width=5cm]{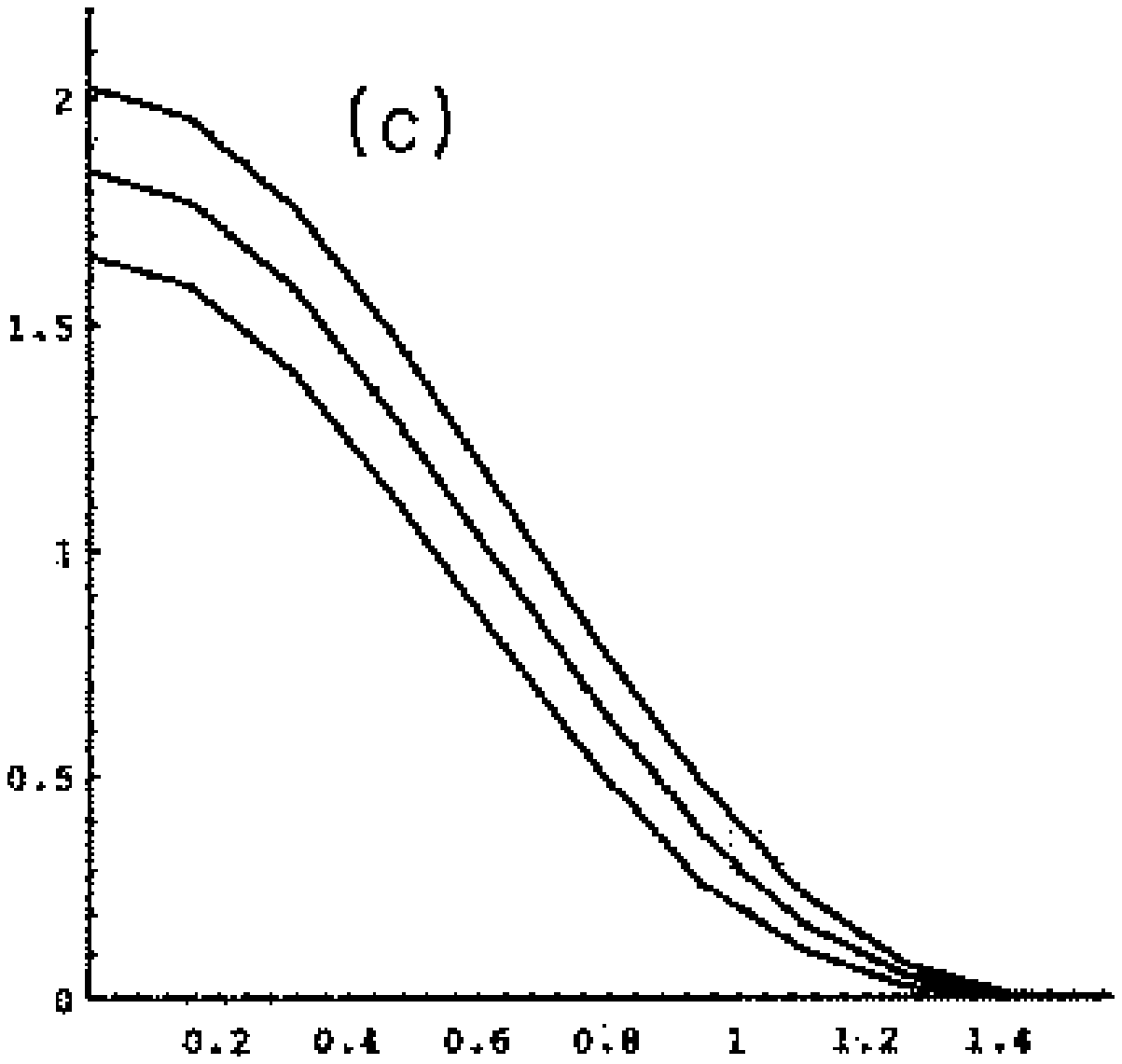}
\caption{The magnitude of the vacuum polarization
$16\pi\langle
T^\mu_\nu\rangle_{twisted}(\rho)/\lambda^{3/2}$ as a function of $\rho$
when the BH mass $M=\pi^{-2}$. Three diagonal components are shown in
separate figures: (a) for $(\mu, \nu)=(t, t)$, (b) $(\mu, \nu)=(r, r)$,
and (c) $(\mu, \nu)=(\theta, \theta)$. The curves correspond to
$\alpha=-1, 0, 1$ as indicated.}
\label{f7}
\end{center}
\end{figure}
\begin{figure}[ht]
\begin{center}
\includegraphics[width=5cm]{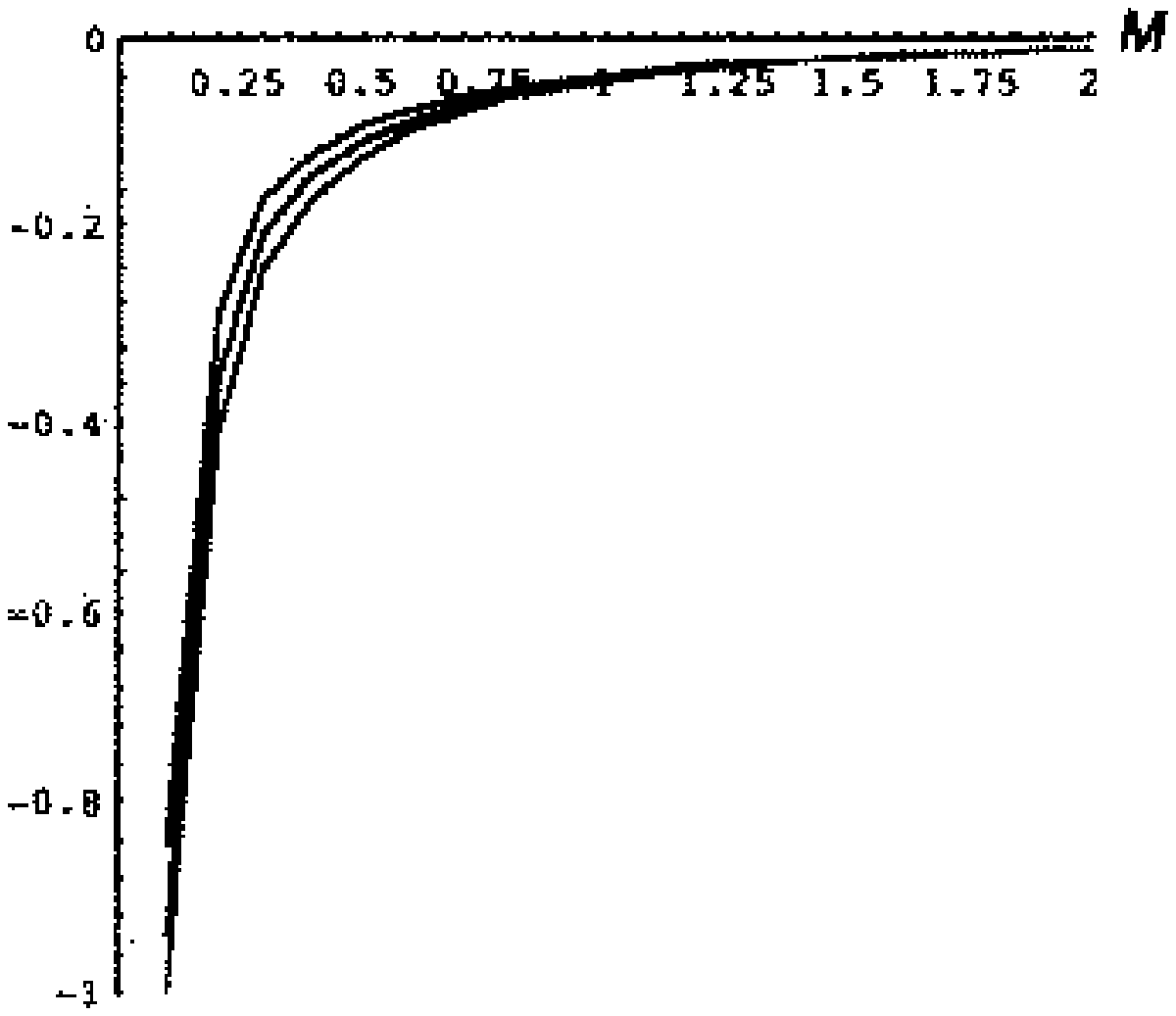}
\caption{The dependence of the vacuum polarization
$16\pi\langle
T^\mu_\nu\rangle_{twisted}(\rho)/\lambda^{3/2}$ on the horizon as a
function of $M$. The curves correspond to $\alpha=-1, 0,
1$ as indicated.}
\label{f8}
\end{center}
\end{figure}

The numerical results for the vacuum stress energy is
shown in Figs.~5--8. The value of $16\pi\langle
T^\mu_\nu\rangle(\rho)/\lambda^{3/2}$ is
plotted in Fig.~5 for $\alpha=-1, 0$, and $1$ when the BH mass
$M=\pi^{-2}$.

The dependence of $\langle
T^\mu_\nu\rangle$ at the horizon $(\rho=0; r=r_H)$
on the BH mass $M$ is shown in Fig.~6. The values of
$\langle
T^\mu_\nu\rangle$ at the horizon approaches zero in the large mass
limit. On the other hand, in the limit of $M\rightarrow 0$, all
components of $\langle
T^\mu_\nu\rangle$ diverge. Note also that $\langle
T^t_t\rangle=\langle
T^r_r\rangle$ is
satisfied on the horizon.

In the limit of $\rho=\pi/2$ ($r\rightarrow\infty$), all the components
of $\langle
T^\mu_\nu\rangle$ for any case become zero. One can regard that
this is due to the ``redshift'' factor $(|g_{tt}|)^{-1/2}$
which goes
to zero at spatial infinity. Thus, thermal equilibrium is
not attained between the BH and the radiation in this system.
For the twisted scalar field, $16\pi\langle
T^\mu_\nu\rangle_{twisted}(\rho)/\lambda^{3/2}$
is exhibited in the same format as the untwisted case in
Figs.~7 and 8. It turns out that the value of $\langle
T^\mu_\nu\rangle$ is not
so sensitive to the value of $\alpha$ for all cases. In the subsequent
section, we consider back reaction of the quantum
effect to the metric.

\section{BACK REACTION TO THE SPACE-TIME METRIC}
In this section, we investigate the effect of back reaction
due to the quantum effects. The back reaction of
quantum fields to the metric has been studied in the
Schwarzschild metric in four dimensions \cite{15,16,17,18}. We
will examine the correction to the gravitational force on
test particles using the corrected metric.

The vacuum values obtained in the preceding section
are quantities of $O(h)$, or one-loop quantum corrections.
The Einstein equation is modified by the vacuum stress
tensor and then becomes
\begin{equation}
R^\mu_\nu-\frac{1}{2}R\delta^\mu_\nu-\lambda\delta^\mu_\nu=\pi\langle
T^\mu_\nu\rangle\,.
\label{4.1}
\end{equation}

We choose the small parameter which denotes the perturbation.
The perturbation parameter, $\varepsilon$, should be proportional
to the Planck constant, which is set by unity
here. Hence we will take $\varepsilon=1/r_H$, since the Planck
length is proportional to h in three dimensions.

Consequently, we need only the expression of the vacuum
stress tensor in the large mass limit. We have for
$\alpha\ne 1$,
\begin{equation}
\langle T^\mu_\nu\rangle_\pm=\pm\frac{(\sqrt{\lambda}r_H)^{3}}{16\pi
r^3}(1-\alpha)e^{-\pi\sqrt{M}}\mbox{diag}(1, 1, -2)+O(e^{-3\pi\sqrt{M}})
\quad
(\alpha\ne 1)
\label{4.2}
\end{equation}
and for $\alpha=1$,
\begin{eqnarray}
& &\langle T^\mu_\nu\rangle_\pm=\pm\frac{(\sqrt{\lambda}r_H)^{3}}{8\pi
r^3}e^{-3\pi\sqrt{M}}\mbox{diag}\left(\frac{9r_H^2}{r^2}-1,
\frac{3r_H^2}{r^2}+4, -\frac{12r_H^2}{r^2}-2\right)\nonumber \\
& &\qquad\qquad+O(e^{-5\pi\sqrt{M}})
\quad
(\alpha=1)\,,
\label{4.3}
\end{eqnarray}
where $\langle T^\mu_\nu\rangle_+$ represents $\langle
T^\mu_\nu\rangle$ for the untwisted scalar and $\langle
T^\mu_\nu\rangle_-$ represents $\langle T^\mu_\nu\rangle_{twisted}$ for
the twisted scalar and the signs on the right-hand side correspond to
them respectively.

We use their leading terms in the analysis in this section.
Of course, they are traceless and covariantly conserved.

We consider the following metric which suffers the
quantum effect:
\begin{equation}
ds^2=-e^{2\psi(r)}[\lambda r^2-m(r)]dt^2+
\frac{dr^2}{\lambda r^2-m(r)}+r^2 d\theta^2
\label{4.4}
\end{equation}
where $m(r)$ and $\psi(r)$ is the function of $r$ which are to be
solved to satisfy the Einstein equations with quantum
correction.

Using the metric (\ref{4.4}), the $(t, t)$ component of the
Einstein equation (\ref{4.1} reads
\begin{equation}
-\frac{1}{2r}\frac{dm(r)}{dr}=\pi\langle T^t_t\rangle\,,
\label{4.5}
\end{equation}
and similarly a linear combination of the $(t, t)$ and
$(r, r)$ components of the Einstein equation leads to
\begin{equation}
-\frac{\lambda r^2-m(r)}{r}\frac{d\psi(r)}{dr}=\pi(\langle
T^t_t\rangle-\langle
T^r_r\rangle)\,.
\label{4.6}
\end{equation}

Equation (\ref{4.5}) is solved in the form
\begin{equation}
m(r)=M[1+\varepsilon\mu(r)]\,,
\label{4.7}
\end{equation}
with
\begin{equation}
\mu(r)=-\frac{2\pi r_H}{M}\int_{r_H}^r\langle
T^t_t\rangle
\bar{r}d\bar{r}\,,
\label{4.8}
\end{equation}
where the mass of the BH is ``renormalized'' so that
$m(r_H)=M$ \cite{15,16,17}.

Similarly, Eq.~(\ref{4.6}) can be solved to the first-order perturbation
in the form 
\begin{equation}
\psi(r)=\varepsilon[\rho(r)-\rho(r_0)]\,,
\label{4.9}
\end{equation}
with
\begin{equation}
\rho(r)=-\pi r_H\int_{r_H}^r\frac{\langle
T^t_t\rangle-\langle
T^r_r\rangle}{\lambda\bar{r}^2-M}\bar{r}d\bar{r}\,,
\label{4.10}
\end{equation}
where we assume the boundary condition $\psi(r_0) =0$
\cite{15,16,17}.

We treat two cases for $\alpha\ne 1$ and for $\alpha=1$ separately.

(1) For $\alpha\ne 1$. We use the leading contribution in (\ref{4.2}).
Then Eq.~(\ref{4.6}) is solved as $\psi=0$, and we get
\begin{equation}
\mu_\pm(r)=\pm\frac{(1-\alpha)(\sqrt{\lambda}r_H)^3}{8M}
e^{-\pi\sqrt{M}}\left(\frac{r_H}{r}-1\right)\quad
(\alpha\ne 1)\,,
\label{4.11}
\end{equation}
Here the plus (minus) sign corresponds to the case with
the untwisted (twisted) scalar field.

(2) For $\alpha=1$. The correction to the BH mass is given,
for a large value of $M$, by
\begin{equation}
\mu_\pm(r)=\pm\frac{(\sqrt{\lambda}r_H)^3}{4M}
e^{-3\pi\sqrt{M}}\left[3\left(\frac{r_H}{r}\right)^3-2\frac{r_H}{r}-1\right]\quad
(\alpha=1)\,,
\label{4.12}
\end{equation}
while substitution of (\ref{4.3}) into (\ref{4.10}) gives
\begin{equation}
\rho_\pm(r)=\pm\frac{(\sqrt{\lambda}r_H)^3}{4M}
e^{-3\pi\sqrt{M}}\left[1-\left(\frac{r_H}{r}\right)^3
\right]\quad
(\alpha=1)\,.
\label{4.13}
\end{equation}

The mass correction $\mu$ take a finite value in the large-$r$
limit in each case. That is
\begin{eqnarray}
m_\pm(\infty)&=&M\mp\varepsilon\frac{(1-\alpha)M^{3/2}}{8}
e^{-\pi\sqrt{M}}\quad(\alpha\ne 1)\,,\label{4.14}\\
m_\pm(\infty)&=&M\mp\varepsilon\frac{M^{3/2}}{4}
e^{-3\pi\sqrt{M}}\qquad\qquad(\alpha=1)\,,
\label{4.15}
\end{eqnarray}
where we leave the parameter $\varepsilon$ which manifestly displays
the perturbation. For the untwisted scalar field, the mass
correction is negative (positive) for $\alpha\le 1$ ($\alpha>1$) in the
asymptotic region. For the twisted scalar field, the sign is
reversed.

Now let us examine the property of the corrected
geometry using the results obtained above. The magnitude
of the static gravitational force on a test particle
placed in the BH geometry is indicated by the ``radial acceleration
in the proper rest frame of the particle''
defined by \cite{16}
\begin{equation}
a=\frac{1}{2}\sqrt{g^{rr}}\frac{\partial}{\partial r}\ln
|g_{tt}|\,.
\label{4.16}
\end{equation}
The positive value for a means the attractive force.

In this expression, we will use the corrected metric up
to the first order in c; the relevant components are
\begin{equation}
g^{rr}(A\lambda r^2-M)\left[1-\varepsilon\frac{M}{\lambda r^2-M}
\mu(r)\right]
\label{4.17}
\end{equation}
and
\begin{equation}
|g_{tt}|(r)=(\lambda r^2-M)\left\{1+\varepsilon\left[2[\rho(r)
-\rho(r_0)]-\frac{M}{\lambda r^2-M}\mu(r)\right]\right\}\,.
\label{4.18}
\end{equation}

Inserting these components into (\ref{4.16}), we obtain
\begin{equation}
a=\frac{\lambda r}{\sqrt{\lambda r^2-M}}\{1+\varepsilon\Delta(r)\}\,,
\label{4.19}
\end{equation}
where the correction term is given by
\begin{equation}
\Delta(r)=\frac{M}{2(\lambda r^2-M)}\mu(r)+
\frac{\pi\sqrt{M}}{\lambda^{3/2}}\langle T^r_r\rangle(r)\,. 
\label{4.20}
\end{equation}

We calculate this using the previous results for the
quantum effect of a conformally invariant scalar field.
For $a\ne 1$, the correction is written in the form
\begin{equation}
\Delta_\pm(r)=\pm(1-\alpha)\frac{\sqrt{M}}{16}e^{-\pi\sqrt{M}}
\frac{\cos^2\rho+\cos\rho-1}{1+\cos\rho}\cos^3\rho
\quad(\alpha\ne 1)\,,
\label{4.21}
\end{equation}
while for $\alpha=1$,
\begin{equation}
\Delta_\pm(r)=\pm\frac{\sqrt{M}}{8}e^{-\pi\sqrt{M}}
\frac{3\cos^3\rho+4\cos^2\rho+\cos\rho-1}{1+\cos\rho}\cos^2\rho
\quad(\alpha=1)\,,
\label{4.22}
\end{equation}
where $\cos\rho=r_H/r$.

We find that, for the untwisted scalar field and $\alpha\le 1$,
the quantum correction gives positive contribution in a
certain region $r<r_c$, and negative contribution in the
outside region. Namely, the quantum effect diminishes
the attractive force in the region $r>r_c$.

The value for $r_c$ is given by
\begin{eqnarray}
& &r_c\simeq 1.618 r_H \quad\mbox{for } \alpha\ne 1\,,
\label{4.23}\\
& &r_c\simeq 2.806 r_H \quad\mbox{for } \alpha=1\,.
\label{4.24}
\end{eqnarray}
For the twisted field or for $\alpha>$1, the sign of the correction
$\Delta$ is reversed.

The correction force is negligibly small for large $M$ and
vanishes at spatial infinity. Therefore, the quantum
correction due to the conformal scalar field cannot be
comparable to the classical attractive force.

\section{DISCUSSION}
In this paper, we have obtained the propagator for a
conformally coupled massless scalar field in 3DBH
space-time with Euclidean signature. Using the exact
propagator, we have computed the vacuum expectation
value $\langle\varphi^2\rangle$ and stress tensor for untwisted and
twisted scalar fields.

The back reaction to the metric has also been discussed
in the present paper. The quantum correction to the
gravitational force on the static test particle has been estimated.
It is found that there are a critical radius outside
the BH, which divides the regions where the positive
and negative correction arise.

For large values of the BH mass, the quantum effects
become exponentially small. One may expect that the
back reaction becomes important only if the BH is small,
or, at the final stage of the BH evaporation. In such
cases, the perturbative analysis is no longer valid.

One may wish to study the thermodynamics of the
3DBH including the quantum field back reaction. Such
analyses have been carried out in Ref.~\cite{17} for four-dimensional
BH's. The approach used in Ref.~\cite{17} is
equally effective for the three-dimensional case. However,
there exist some subtleties in the 3DBH case which
arise from the existence of the additional length scale
$\lambda^{-1/2}$ and the $h$ independence of the Planck mass in three
dimensions. We leave the discussion on the thermodynamics
with quantum fields for separate publications.

In the present paper, we discussed only conformal
massless scalar field. The other types of the fields should
be taken into the general analysis.

The quantum-field theory around a rotating 3DBH is
an interesting subject worth studying. The thermodynamics
of the rotating 3DBH's including quantum
fields should be clearly understood in the future.

\bigskip

\noindent
\textit{Note added.} After completion of this manuscript, we
became aware of the works by Steif \cite{19} and by Lifshifts
and Ortiz \cite{20}. Steif calculated the stress tensor
$\langle T^\mu_\nu\rangle$ for the scalar field with transparent
boundary condition ($\alpha=0$ in our case). He also studied the
rotating BH background. Lifschytz and Ortiz considered the Dirichlet
and Neumann boundary conditions ($\alpha=1$ and $-1$
in our case, respectively). They also calculated a response
function for a scalar field outside the BH horizon.

\section*{ACKNOWLEDGMENT}
This work was supported in part by a Grant-in-Aid for
Scientific Research from the Ministry of Education, Science
and Culture (No. 05740186).



\begin{thebibliography}{99}
\bibitem{1} \textit{Quantum Theory of Gravity}, edited by S. M.
Christensen (Hilger, Bristol, 1984); \textit{Conceptual Problems of
Quantum gravity}, edited by A. Ashtekar and J. Stachel (Birkhauser,
Boston, 1991); E. Alvarez, Rev. Mod. Phys. {\bf 61} (1989) 561.
\bibitem{2} N. D. Birrell and P. C. W. Davies, \textit{Quantum Fields in
Curved Space} (Cambridge University Press, Cambridge,
England, 1982).
\bibitem{3} P. Candelas, Phys. Rev. {\bf D21} (1980) 2185; P. Candelas
and K. W. Howard, ibid. {\bf D29} (1984) 1618; K. W. Howard
and P. Candelas, Phys. Rev. Lett. {\bf 53} (1984) 403; K. W.
Howard, Phys. Rev. {\bf D30} (1984) 2532; V. P. Frolov, ibid.
{\bf D26} (1982) 954; V. P. Frolov, F. D. Mazzitelli and J. P.
Paz, ibid. {\bf D40} (1990) 948; I. D. Novikov and V. P. Frolov,
\textit{Physics of Black Holes} (Kluwer Academic, Dordrecht,
1988); V. P. Frolov, in \textit{Trends in Theoretical Physics}, edited
by P. J. Ellis and Y. C. Tang (Addison-Wesley, Reading,
MA, 1991), Vol. 2, pp. 27--75; P. R. Anderson, Phys.
Rev. {\bf D39} (1989) 3785; {\bf D41} (1990) 1152.
\bibitem{4} M. Banados, C. Teitelboim and J. Zanelli, Phys. Rev.
Lett. {\bf 69} (1992) 1849; M. Bandados, M. Henneaux, C.
Teitelboim and J. Zanelli, Phys. Rev. {\bf D48} (1993) 1506.
\bibitem{5} S. F. Ross aud R. B. Mann, Phys. Rev. {\bf D47} (1993)
3319; D. Cangemi, M. Leblanc and R. B. Mann, ibid. {\bf D48} (1993)
3606; A. Achucarro and M. Ortiz, ibid. {\bf D48} (1993) 3600; G. T.
Horowitz and D. L. Welch, Phys. Rev. Lett. {\bf 71} (1993) 328; N.
Kaloper, Phys. Rev. {\bf D48} (1993) 2598; C. Farina, J. Gamboa and A.
J. Segui-Santonja, Class. Quantum Grav. {\bf 10} (1993) L193.
\bibitem{6} K. Shiraishi and T. Maki, Class. Quantum Grav.
{\bf 11} (1994) 695.
\bibitem{7} C. W. Misner, K. S. Thorne and J. A. Wheeler,
\textit{Gravitation} (Freeman, San Francisco, 1973).
\bibitem{8} S. J. Avis, C. J. Isham and D. Storey, Phys. Rev. {\bf D18}
(1978) 3565.
\bibitem{9} C. P. Burgess and C. A. Lutken, Phys. Lett. {\bf B153}
(1985) 137.
\bibitem{10} C. J. C. Burges et al., Ann. Phys. (N.Y.) {\bf 167} (1986)
285.
\bibitem{11} B. Allen, A. Folacci and G. W. Gibbons, Phys. Lett.
{\bf B189} (1987) 304.
\bibitem{12} A. G. Smith, in \textit{The Formation and Evolution of
Cosmic Strings}, edited by G. W. Gibbons, S. W. Hawking, and T.
Vachaspati (Cambridge University Press, Cambridge,
England, 1990).
\bibitem{13} For an example, V. P. Frolov, C. M. Massacand and C.
Schmid, Phys. Lett. {\bf B279} (1992) 29.
\bibitem{14} C. J. Isham, Proc. R. Soc. London {\bf A362} (1978) 383;
{\bf A364} (1978) 591; L. H. Ford, Phys. Rev. {\bf D21} (1980) 949.
\bibitem{15} J. W. York, Phys. Rev. {\bf D31} (1985) 775.
\bibitem{16} D. Hochberg and T. W. Kephart, Phys. Rev. {\bf D47} (1993)
1465; D. Hochberg, T. W. Kephart and J. W. York, Phys. Rev. {\bf D49}
(1994) 5257.
\bibitem{17} D. Hochberg, T. W. Kephart aud J. W. York, Phys. Rev.
{\bf D48} (1993) 479.
\bibitem{18} J. W. York, Phys. Rev. {\bf D33} (1986) 2092.
\bibitem{19} A. Steif, Phys. Rev. {\bf D49} (1994) 585.
\bibitem{20} G. Lifschytz aud M. Ortiz, Phys. Rev. {\bf D49} (1994)
1929.
\end{thebibliography}
\end{document}